\documentclass[prl,twocolumn]{revtex4}
\usepackage{graphicx}
\usepackage{color}
\usepackage{amsmath}
\usepackage{amssymb}
\usepackage{latexsym}

\begin{document}
\title{Spin splitting in open quantum dots}

\author{M. Evaldsson, I. V. Zozoulenko}
\affiliation{ Department of Science and Technology (ITN), Link\"{o}ping University, 601\,74
Norrk\"oping, Sweden}
\author{M. Ciorga, P. Zawadzki, A. S. Sachrajda}
\affiliation{Institute for Microstructural Science, National Research Council, K1A 0R6, Ottawa,
Canada}

\date{\today}

\begin{abstract}
We demonstrate that the magnetoconductance of small lateral
quantum dots in the strongly-coupled regime (i.e. when the leads
can support one or more propagating modes) shows a pronounced
splitting of the conductance peaks and dips which persists over a
wide range of magnetic fields (from zero field to the edge-state
regime) and is virtually independent of the magnetic field strength. Our
numerical analysis of the conductance based on the Hubbard
Hamiltonian demonstrates that this is essentially a many-body/spin
effect that can be traced to a splitting of degenerate levels in
the corresponding closed dot. The above effect in open dots can be
regarded as a counterpart of the Coulomb blockade effect in weakly
coupled dots, with the difference, however, that the splitting of
the peaks originates from the interaction between the electrons of
opposite spin.
\end{abstract}
\maketitle

There has been a lot of interest recently in the spin properties
of semiconductor quantum dots. This is due not only to the new
fundamental physics that these devices exhibit but also to
possible applications in the emerging fields of
spintronics and quantum information.
%Small symmetric dots exhibit properties reminiscent of those of
%real atoms, including the formation of a shell structure filled
%according to Hunds's rule that favors a ground state with maximum
%possible spin \cite{Tarucha,Hawrylak1996,Reimann}.
%Recent theoretical studies also predict that even in the case of
%large disordered or chaotic quantum dots, interactions can lift
%the spin degeneracy and lead to a spontaneous ground state
%magnetization \cite{Berkovits,Andreev,Brouwer}. However, no clear
%consensus has been reached so far on whether states with minimum
%or non-minimum spins dominate in large quantum dots
%\cite{Stone,Wingreen,Sivan,Vallejos,Luscher}.
Coulomb blockade (CB) experiments are often used to probe the
nature of the spin states
\cite{Tarucha,Sivan,Tarucha2000,Luscher,MariuszPRB,HawrylakPRB,CiorgaPRL,Fuhrer}.
The CB regime corresponds to a weak coupling between the dot and
the leads, so that the number of electrons in the dot is integer
and each peak signals an addition/removal of one electron to/from
the dot. In strong contrast to the Coulomb blockade regime in the
\textit{open dot} dot regime electrons can freely enter and exit
the dot via leads that support one or more propagating modes. In
this case the charge quantization no longer holds and one may
expect that the conductance is mediated by two independent
channels of opposite spin resulting in a total spin $S=0$ in the
dot. The degree of spin degeneracy in this regime was probed in
Ref. \cite{Marcus} for a large chaotic dot where the statistical
analysis of the conductance fluctuations indicated that a dot was
spin-degenerate at low magnetic fields. In the present paper we
present experimental evidence that in small open dots two spin
channels are correlated and therefore \textit{the spin degeneracy
can be lifted}.

\begin{figure}[!htp]
\includegraphics[scale=1.1]{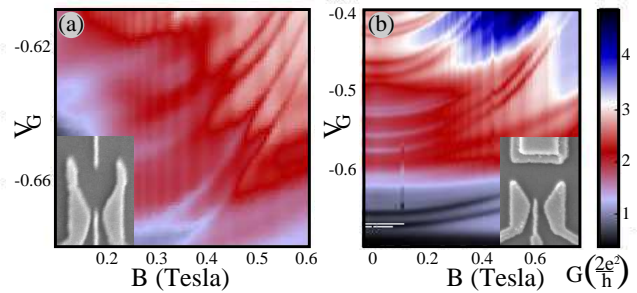}
\caption{Experimental conductance as a function of magnetic field
$B$ and gate voltage $V_G$ obtained from different dots. The
corresponding device layouts are shown in the insets. The
lithographic size of the dots is $\sim 450$nm.} \label{fig1}
\end{figure}

A gate device layout scheme has been recently developed which
enables the number of electrons confined within an
electrostatically defined quantum dot to be controllably reduced
to zero \cite{MariuszPRB}. These few electron devices were used to
study the spin properties of quantum dots in the CB regime using
Coulomb and spin blockade spectroscopic techniques
\cite{HawrylakPRB,CiorgaPRL}. The measurements in this paper are
on these same devices but in the strongly coupled regime using a
resistance bridge with $\sim$1nA current with an estimated number
of electrons in the dot from 25 to 90. Details of the two device
designs and the AlGaAs/GaAS wafer used for the measurements are
given elsewhere \cite{HawrylakPRB,CiorgaPRL}. For the open dot
experiments the following experimental procedure was used. All the
gates defining the quantum dot were swept simultaneously. The
ranges of the sweeps on the individual gates were not identical
but were chosen, making use of calibration measurements, to
maintain approximately the same conductance at both the entrance
and exit leads (i.e. the voltage was applied to the gates in a way
at any point on the trace the entrance and exit QPCs contained the
same number of propagating modes). The voltage in the figure
corresponds to the value applied to one ``representative" gate --
the other voltages would be similar but not identical. Altogether
measurements were made on four different quantum dots. Figure
\ref{fig1} illustrates typical experimental results. As can be
seen clearly in the data there exists a remarkable splitting of
all of the conductance peaks. The features discussed in this paper
were present in all four dots and on several cooldowns. The amount
of splitting varies from doublet to doublet with a typical value
of 0.2meV. We stress that the observed splitting is almost two
orders of magnitude larger that Zeeman splitting (which is
~0.005meV at B=0.5T). The splitting of the conductance peaks is
the main experimental result of this paper.

%____________________________________________________________________________________
%____________________________________________________________________________________

In order to model the magnetoconductance through the dot we use
a tight-binding Hubbard Hamiltonian
%where many-particle and spin effects are included
in the mean-field approximation
\cite{PWAnderson,Lee,Henrickson,Berkovits,Sivan,Nonoyama},
$H=H_\uparrow +H_\downarrow$,

\begin{eqnarray}\label{Hamiltonian}
H_\sigma=
%\sum_{\mathbf{r}}\left(\epsilon_0+V_\mathbf{r}\right)
%a^\dag_{\mathbf{r}\sigma}a_{\mathbf{r}\sigma}
-\sum_{\mathbf{r,\Delta}}
%\left
t_{\mathbf{r},\Delta} a^\dag_{\mathbf{r}\sigma}
a_{\mathbf{r+\Delta}\sigma }
%\right)
+ U\sum_{\mathbf{r}} \langle a^\dag_{\mathbf{r}\sigma'}
a_{\mathbf{r}\sigma'}\rangle  a^\dag_{\mathbf{r}\sigma}
a_{\mathbf{r}\sigma},
\end{eqnarray}
where $\sigma, \sigma'$ describe two opposite spin states
$\uparrow,\downarrow$ (or $\uparrow,\downarrow$),
$a^\dag_{\mathbf{r}\sigma}$ and $a_{\mathbf{r}\sigma}$ are the
creation and annihilation operators at the lattice cite
$\mathbf{r}$ for an electron with spin $\sigma$, $\mathbf{\Delta}$
corresponds to nearest neighbors, and $t_{\mathbf{r},\Delta}$ is a
hopping matrix between the neighboring sites where the magnetic
field $B$ is incorporated in a standard fashion as the phase
factor in the form of Peierls' substitution \cite{Lee}, and the
onsite Hubbard constant $U$ describes the Coulomb interaction
between electrons of different spin. The Hamiltonian
(\ref{Hamiltonian}) does not include the classical charging term
$e^2/C$ because we concentrate on the open dot regime when the
charging in unimportant. Note that the actual dot potential is
unknown. On the basis of the experimental findings we however
expect that the observed effect of the spin-splitting is generic
to small quantum dots and thus rather insensitive to a detailed
shape of the potential. We thus use the hard-wall confinement for
the dot of a rectangular shape with the size $0.21\times 0.36
\mu$m that is connected to infinite ideal leads with width $w=80$
nm (see below, inset to Fig. \ref{fig3}). We note that because of
the uncertainty of the actual potential profile we do not expect a
one-to-one correspondence between the calculated and measured
conductance. The conductance of the dot is given by the Landauer
formula $G =G_\uparrow + G_\downarrow
=\frac{e^2}{h}\left(T_\uparrow + T_\downarrow\right)$, where
$T_\sigma$ is the transmission coefficient for different spin
channels. In order to calculate $T_\sigma$ we introduce the
retarded Green function ${\cal G}_\sigma=(E-H_\sigma+i\epsilon)$
and employ the standard recursive Green function technique
\cite{Lee,Datta}. The expectation value for the electron number at
site $\mathbf{r}$ for the spin $\sigma$ is given by \cite{Datta}
\begin{eqnarray}\label{<n>}
\langle N_{\mathbf{r}\sigma}\rangle=\langle
a^\dag_{\mathbf{r}\sigma} a_{\mathbf{r}\sigma}\rangle=
-\frac{1}{\pi}\int_{-\infty}^{E_F}\mathrm{Im}\left[{\cal
G}_\sigma(\mathbf{r},\mathbf{r},E)\right]dE ,
\end{eqnarray}
where $E_F$ is the Fermi energy, and ${\cal
G}_\sigma(\mathbf{r},\mathbf{r},E)$ is the Green function in the
real space representation. Equations (\ref{Hamiltonian}),
(\ref{<n>}) are solved self-consistently. Because all the poles of
the Green function are in the lower complex plane, the integration
path in Eq. (\ref{<n>}) can be transformed into the upper complex
plane where the Green's function is smoother than on the real
axis. This also allows us to account for the bound states in the
dot that are situated below the propagation threshold in the
leads. In the calculations the lattice constant is chosen to be
$a=10$ nm that insures that Eq. (\ref{Hamiltonian}) with
$|t|=\frac{\hbar^2}{2m^*a^2}$ corresponds to a continuous
Schr\"odinger equation, with $m^*=0.067m_0$ being the effective
electron mass appropriate for GaAs. We neglect the Zeeman term as
its effect on the dot conductance in the chosen field interval is
negligible. All the results presented in the paper correspond to a
typical value of $U=3|t|$ \cite{Berkovits,Henrickson}. (We also
performed calculations where $U$ was varied in a broad range
$t<U<7t$ and arrived to the qualitatively same results). Note that
the present Hamiltonian reproduces Hund's rule for the
eigenspectrum of a closed parabolic dot
\cite{Tarucha,Hawrylak1996}.

%\cleardoublepage
%-------------------------- Fig 2--------------------------
%\begin{figure*}
\begin{figure}
\includegraphics[width=0.45\textwidth]{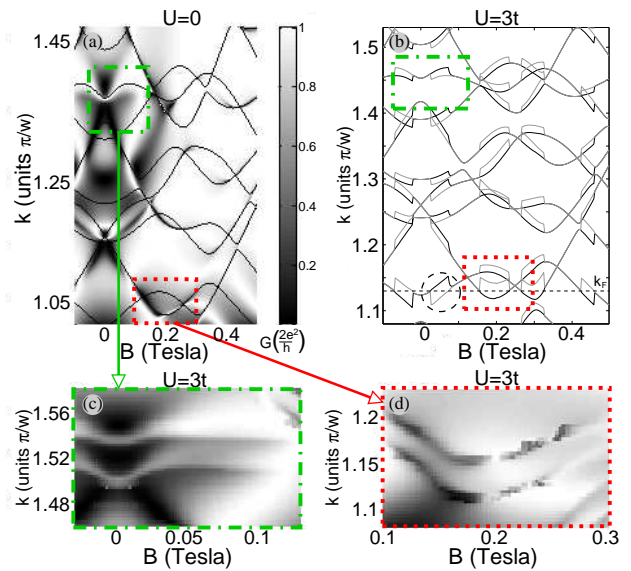}
 \caption{\label{fig2} (a) The conductance $G=G(k_F,B)$  for $U=0$.
 Solid lines depict the
eigenspectrum of the corresponding closed dot.
%The number of modesin the leads is $M=\textrm{int} [kw/\pi]$.
(b) The eigenspectrum of the corresponding closed dot for the same
strength of $U=3t$. Dotted line indicates $k_F$.(c),(d) The
conductance $G=G(B,k_F)$ for $U=3t$. Note that because of a large
computation time the conductance is shown only for two
representative regions.
%The rectangular in the panel (a) indicates a corresponding region
%in the linear plot ($U=0$), where the present region transforms
%from.
 }
\end{figure}
%\end{figure*}
%----------------------------------------------------------

Figure \ref{fig2} (a) shows a linear conductance vs magnetic field
$B$ and Fermi wave vector $k_F$ which includes no spin or Coulomb
effects ($U=0$). For a comparison the single-particle spectrum of
the corresponding closed dot is superimposed onto conductance plot
in order to underline the relationships between them. The
eigenspectrum of the dot for the case of $U=3|t|$ is shown in Fig.
\ref{fig2} (b) for a representative $k_F$ in the dot. The
principle features of the single-particle eigenspectrum can still
be traced in the eigenspectrum of Hubbard Hamiltonian
(\ref{Hamiltonian}). The spectrum of Eq. (\ref{Hamiltonian}) is
however shifted to higher energies because of the increased dot
electrostatic potential due to the charge build-up described by
the second term in Eq. (\ref{Hamiltonian}). The major difference
in comparison with the single-particle spectrum is that for
certain regions of magnetic field the spin degeneracy is lifted
and thus spin-up and spin-down eigenenergies are split. We shall
demonstrate below that spin splitting effect is directly related
to the Hubbard term in the Hamiltonian, where the spin species of
one sort feel the potential from the electrons of the opposite
spin. (This effect is absent for a spinless Hamiltonian with the
Hartree-type term alone, as the spin-up and spin-down electrons
would feel the same potential).
 Suppose, a particular
eigenstate is spin-degenerate, close to $E_F$, and occupied by two
electrons of opposite spin (let us concentrate for example on the
eigenlevel that is close to the $E_F$ in the interval marked by a
dashed circle in Fig. \ref{fig2} (b)). As the magnetic field
increases, the eigenenergy increases and eventually crosses $E_F$.
If the Hamiltonian were spinless, this eigenstate would become
totally unpopulated. However, because of the Hubbard term in Eq.
(\ref{Hamiltonian}), spin-up and spin-down electrons feel the
Coulomb potential from electrons with opposite spin. Thus, if only
one of the electrons, say spin-up, remains in the dot, it does not
feel the electrostatic potential from the spin-down fellow
electron it was sharing the same eigenlevel with, because the
latter is no longer in the dot. As a result, the effective
potential for the spin-up electron drops abruptly when the
spin-down electron leaves the dot. This results in an abrupt drop
of the eigenvalue for the spin-up electron, hence lifting of the
spin-degeneracy of the spectrum. As the magnetic field increases
further the eigenlevel under consideration (now singly occupied)
rises and eventually crosses $E_F$ (at $B\sim 0.08T$). The level
thus becomes totally depopulated and therefore the effective
potentials for spin-up and spin-down electrons become equal. This,
in turn, leads to the restoration of the spin degeneracy. Note
that the origin of the spin-splitting of the eigenspectrum of the
quantum dot described above is conceptually similar to Anderson
model of the formation of the localized magnetic states on solute
ions in nonmagnetic metals \cite{PWAnderson}.

%-------------------------- Fig 3--------------------------
\begin{figure}
\resizebox{0.9\linewidth}{!}{
\includegraphics[angle=-90]{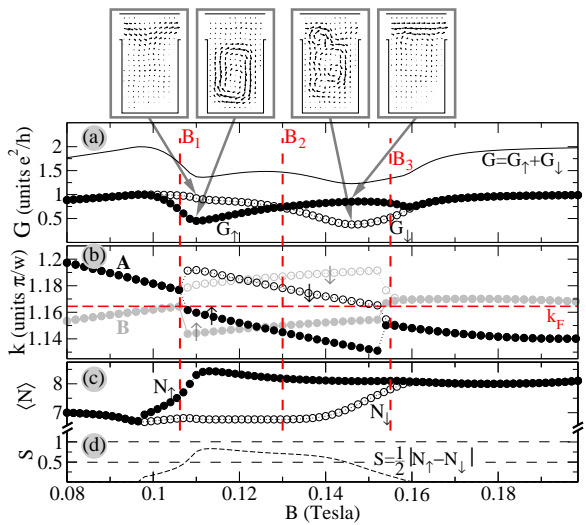}
} \caption{\label{fig3} (a) Conductance of the open quantum dot
$G=G(B)$ for spin-up and spin-down electrons, $G_\uparrow$ and
$G_\downarrow$, respectively. (b) magnetospectrum of the related
\textit{closed} dot; $k_Fw/\pi=1.165$. (c) Number of spin-up and
spin-down electrons in the open quantum dot, $N_\uparrow$ and
$N_\downarrow$, respectively. (c) Total spin of the electrons in
the related open quantum dot, $S=\frac{1}{2}|N_\uparrow -
N_\downarrow|$. Insets illustrate the current density for spin-up
and spin-down electrons in the open dot.}
\end{figure}
%----------------------------------------------------------

\begin{figure}
\resizebox{0.9\linewidth}{!}{
\includegraphics[angle=-90]{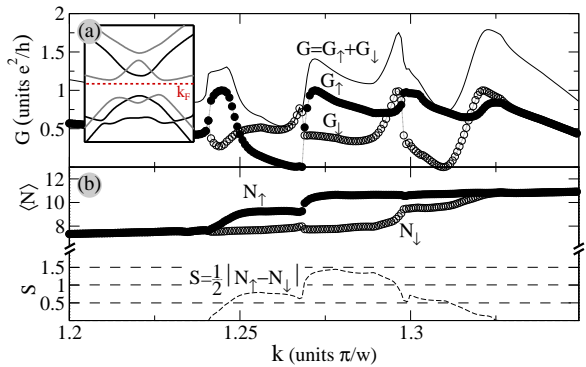}
} \caption{\label{fig4}  Conductance of the quantum dot,
$G_\uparrow$ and $G_\downarrow$ (a), and the number of spin-up and
spin-down electrons, $N_\uparrow$ and $N_\downarrow$ (b),  as a
function of $k_F$ at $B=0$. Inset shows a corresponding
eigenspectrum in the vicinity of $B=0$.}
\end{figure}

The spin splitting in the quantum dot eigenspectrum leads to a
splitting of the conductance peaks/dips which become doublets as
illustrated in Fig. \ref{fig2}(c),(d). (Note that because of
significant computational time, the splitting of the peaks are
shown as a 2D grey scale plots for two selected regions only. All
other peaks/dips in other regions show the similar behavior). A
detailed analysis of the doublet formation and its relation to
splitting of the eigenlevels of the corresponding closed dot is
discussed in Fig. \ref{fig3}. Note that in a closed dot
\textit{all} states in the vicinity of $E_F$ equally contribute to
the spin splitting. In contrast, in an open dot, the states
strongly coupled to the leads (with wide resonant broadenings
$\Gamma$) have very little effect on the spin splitting. This is
because of a short lifetime of these states $\tau\sim\hbar/\Gamma$
which is not long enough to provide a sufficient charge build-up
in the dot. In the $B$-field interval of Fig. \ref{fig3}, there
are two eigenstates in the vicinity of $E_F$, labelled as $A$ and
$B$. An analysis of the eigenfunction and linear conductance
($U=0$) shows that the state $B$ is strongly coupled to the leads
and thus provide a nonresonant channel of transport with $T\approx
1$. In our further analysis we therefore will concentrate only on
the state $A$ which is weakly coupled to the leads and thus
responsible for a resonant channel of the conductance and the
splitting of the peaks. When the magnetic field $B\lesssim B_1$,
this state is empty, see Fig. \ref{fig3} (b). When this eigenstate
approaches $E_F$ it splits because of the reasons discussed in the
preceding paragraph. As the transport through the dot at zero
temperature occurs at $E_F$, the spin-up resonant state will
affect the conductance more strongly than the spin-down resonant
state. This is because the former is situated at $E\approx E_F$,
whereas the later is shifted from $E_F$ by the distance determined
by effective electrostatic potential from the spin-up electrons,
see Fig. \ref{fig3} (a),(b). Therefore, the first dip in the
doublet is caused by spin-up electrons. (Note that unlike the CB
regime the resonant state in an open dot can give rise to either a
peak or a dip depending on the interference condition). As the
magnetic field increases further, the spin-up resonant state moves
farther away from $E_F$, whereas the spin-down state moves towards
$E_F$. At some field $B_2$ the distances from the resonant states
to $E_F$ becomes equal and both states contribute equally to the
conductance. For $B\gtrsim B_2$ the spin-down resonant state
becomes dominant and thus the second dip in the doublet is due to
spin-down electrons. Eventually, when $B\approx B_3$ both resonant
states become populated and the spin degeneracy is lifted. Insets
in Fig. \ref{fig3} show the current density in the dot
illustrating the role played by the bound states in formation of
dips in the conductance doublet. Figure \ref{fig3} (c) shows the
electron number $N$ and total spin polarization
$S=\frac{1}{2}|N_\uparrow -N_\downarrow|$ inside the open quantum
dot. It demonstrates that neither $N$ nor $N_\uparrow,
N_\downarrow$ are integer. This is in contrast to the case of
weakly coupled dots in the Coulomb blockade regime when $N$ is
always integer. We also find that in the vicinity of doublets $S$
is distinct from 0, (Fig. \ref{fig3} (d), see also below, Fig.
\ref{fig4} (b)). This contradicts the conclusion of Folk
\textit{et al. } \cite{Marcus} that an open dot is
spin-degenerate. This difference may be attributed to the fact
that Folk \textit{et al. } studied relatively large dots, and it
is not clear whether the spin-splitting effect discussed here for
small dots would survive in larger dots. Note that similar effects
of spin splitting have been investigated in a number of model
systems \cite{Nonoyama}. In addition, spontaneous spin splitting
has been suggested as the origin of the ``0.7-anomaly" in the
conductance of a quantum point contact \cite{QPC}.

We would like to point out that the observed effect of the
spin-splitting in open dots can be regarded as a counterpart to
the Coulomb blockade effect in the weakly-coupled (closed) dots,
because both effects are caused by the charging of the dot (Note
that one usually do not expect charging in the open dot regime).
The essential difference is that the CB peaks are separated by the
distance given by a classical charging energy arising from the
addition of one extra electron to the dot (regardless of its
spin), whereas for the case of small open dots the peak splitting
is determined by a charging effect that arises from the interaction of
electrons of opposite spin.

%-------------------------- Fig 4--------------------------
%----------------------------------------------------------

The splitting of the conductance peaks/dips is the most pronounced
manifestation of spin polarization in an open dot. However, the
spin degeneracy in the dot can be lifted even when the conductance
does not show an apparent doublet formation. This is illustrated
in Fig. \ref{fig4} where the dot conductance is plotted in the
region containing three closely spaced eigenstates as shown in the
inset. In this region the spacing between the levels is smaller
than the energy splitting between spin-up and spin-down levels. As
a result, more than one eigenstate can contribute to a particular
peak/dip, and the conductance shows an erratic behavior where it
is not possible to identify well-defined spin-split doublets (Fig.
\ref{fig4} (a)). Nevertheless, in the given energy interval the
electron density shows a pronounced polarization as indicated in
Fig. \ref{fig4} (b).

It is interesting to note that practically all resonant peaks and
dips have a characteristic asymmetric shape as a function of
$k_F$. This is a signature of a Fano resonance that occurs in an
open dot because of an interference between a resonant channel
(related to a spin-resolved resonant state) and a non-resonant one
(originated from the contributions from the tails of neighboring
levels) \cite{Fano}.

In conclusion, we find experimentally that conductance peaks and dips
are split in small few electron open quantum dots. The numerical analysis of the
conductance and the dot eigenspectrum demonstrates that this
effect is related to a spin splitting in the corresponding
closed dot when the interactions between the electrons with
opposite spins is taken into account.

\begin{acknowledgments}
Financial support from the National Graduate School in Scientific
Computing (M.E.) and the Swedish Research Council (M.E. and I.V.Z)
is acknowledged. A.S.S would like to acknowledge support from the Canadian Institute for Advanced Research.
\end{acknowledgments}

\end{document}